\documentclass[aps]{revtex4}
\usepackage{graphicx}
\begin{document}

\title{The Tensor Current Divergence Equation in $U(1)$ 
Gauge Theories is Free of Anomalies}
\author{Wei-Min Sun$^1$, Hong-Shi Zong$^{2,1}$, 
Xiang-Song Chen$^1$ and Fan Wang$^1$}
\affiliation{$^1$Department of Physics 
and Center for Theoretical Physics, Nanjing University, 
Nanjing 210093, China \\
$^2$CCAST(World Laboratory), P.O.Box 8730, Beijing 10080, China}

\begin{abstract}
The possible anomaly of the tensor current divergence equation in 
$U(1)$ gauge theories is calculated by means of perturbative
method. It is found that the tensor current divergence equation 
is free of anomalies.
\end{abstract}
\maketitle

The Dyson-Schwinger equations(DSEs) \cite{Roberts, Alkofer} provide 
a nonperturbative continuum framework for solving gauge field theories. 
The DSEs are an infinite tower of coupled integral equations which 
relate the $n$-point Green function to the $(n+1)$-point function;
at its simplest, propagators are related to three-point vertices and
so on. In order to actually solve the coupled equations one must find
some way to truncate the system. For example, if one specifies the
form of the fermion-boson vertex function, then the DSEs form
a closed system for the propagators and can therefore be solved.
Here the Ward-Takahashi(WT) identity \cite{WardTakahashi}can be used 
to constrain the
form of the fermion-boson vertex function. However, the normal WT 
identity(which may be called longitudinal WT identity) only
specifies the longitudinal part of the vertex function, leaving
its transverse part undetermined. More specifically, the longitudinal
part of the fermion-boson vertex function can be written in terms of
the full fermion propagator as
\begin{equation}
k_\mu \Gamma^{\mu}(q,p)=S^{-1}(q)-S^{-1}(p), k_\mu \equiv q_\mu-p_\mu
\end{equation}
It is evident that the above WT identity does not at all specify the 
transverse part of the vertex. In order to further constain the transverse
part of the vertex function, some authors have studied the so-called 
transverse WT identities \cite{Kondo, He1, He2} which specify the curl of
the vertex. In the transverse WT identities derived in \cite{He1,He2}, 
the vertex functions of different Lorentz structurs are coupled with 
each other. Therefore, in order to constrain the transverse part of 
the fermion-boson vertex function(vector vertex function), it is 
insufficient to consider the transverse WT identity for this vertex
function alone. Instead, one should consider the whole set of  
WT identities(including both longitudinal and transverse ones) for vertex
functions of different Lorentz structures. The longitudinal(transverse) 
WT identities can be derived from the divergence(curl) equation of the 
corresponding fermionic bilinear currents. Due to singularities of product
of local field operators, these divergence(or curl) equations may
suffer from anomalies at the quantum level and correspondingly we have
the anomalous WT identities. A well-known example is the Adler-Bell-Jackiw
(ABJ) anomaly \cite{ABJ} for the divergence equation of the axial vector current.
The possible anomalies for the curl equation of the axial vector and vector 
current in $U(1)$ gauge theories are studied by He \cite{HeAnomaly} and us
\cite{Sun} using the point-splitting method and perturbative method, respectively.
It is found that the axial vector and vector current curl equations are free of
anomalies. In this paper we shall use similar methods to study the possible 
anomaly of the tensor current divegence equation in $U(1)$ gauge theories.

First let us see the classical divergence equation of the tensor current
in a theory of a classical Dirac field interacting with an external $U(1)$
gauge field. The Lagrangian is
\begin{equation}
{\cal L}={\bar \psi}(x)(i \not\! D-m)\psi(x)
\end{equation}
where $D_\mu=\partial_\mu-ie A_\mu(x)$ is the covariant derivative. By using 
the equation of motion for the Dirac field, one can verify that
\begin{eqnarray}
&&\partial_\rho({\bar \psi}(x)\frac{i}{2}(\gamma^{\rho}\Omega-{\tilde \Omega}
\gamma^{\rho})\psi(x)) \nonumber \\
&=&{\bar \psi}(x)\frac{i}{2}(\gamma^{\rho}\Omega+{\tilde \Omega}\gamma^{\rho})
(\stackrel{\rightarrow}{\partial_\rho}-\stackrel{\leftarrow}{\partial_\rho})
\psi(x)+e{\bar \psi}(x)A_\rho(x)(\gamma^{\rho}\Omega+{\tilde \Omega}\gamma^{\rho})
\psi(x) \nonumber \\
&&-m{\bar \psi}(x)(\Omega+{\tilde \Omega})\psi(x) 
\end{eqnarray}
where $\Omega$ and ${\tilde \Omega}$ are two arbitrary $4\times 4$ matrices in spinor
space. Now we choose $\Omega={\tilde \Omega}=\gamma^{\nu}$ in the above equation and 
obtain
\begin{eqnarray}
&& \partial_\mu({\bar \psi}(x)\sigma^{\mu\nu}\psi(x)) \nonumber \\
&=& -2m{\bar \psi}(x)\gamma^{\nu}\psi(x)+i{\bar \psi}(x)(\stackrel{\rightarrow}
{\partial^{\nu}}-\stackrel{\leftarrow}{\partial^{\nu}})\psi(x)+2e{\bar \psi}(x)
A^{\nu}(x)\psi(x) 
\end{eqnarray}
which is the classical divergence equation of the tensor current. Now let us quantize 
the Dirac field(but still treat the $U(1)$ gauge field as a classical external
field), then the above divergence equation may suffer from anomalies due to singularities
of product of local field operators. In the following we shall calculate this possible
anomaly term using perturbative method.

The quantized form of the tensor current divergence equation, including possible anomaly
term, is
\begin{eqnarray}
&&\partial_\mu\langle {\bar \psi}(x)\sigma^{\mu\nu}\psi(x)\rangle \nonumber \\
&=&-2m\langle {\bar \psi}(x)\gamma^{\nu}\psi(x)\rangle+i\langle {\bar \psi}(x)(
\stackrel{\rightarrow}{\partial^{\nu}}-\stackrel{\leftarrow}{\partial^{\nu}})
\psi(x)\rangle+2e\langle{\bar \psi}(x)\psi(x)\rangle A^{\nu}(x) \nonumber \\
&&+anomaly \label{tensoranomaly}
\end{eqnarray}
where $\langle \cdots \rangle$ stands for the vacuum expectation value(VEV) in the
presence of the external $U(1)$ gauge field. In the following we shall calculate 
perturbatively the VEVs appearing in Eq.(\ref{tensoranomaly}) and see whether there
exists an anomaly term. The VEV of each operator appearing in Eq.(\ref{tensoranomaly})
is a functional of the external field $A_\mu$. After expanding it in powers of $A_\mu$
we get a series of one-loop diagrams. Now let us analyse these VEVs order by order
in $A_\mu$.

The $A^n$ order contribution to $\langle {\bar \psi}(x)\sigma^{\mu\nu}\psi(x)\rangle$ 
is represented by an $(n+1)$-gon diagram(see fig.1) and equals
\begin{eqnarray}
\int\frac{d^4 q_1}{(2\pi)^4}\cdots \frac{d^4 q_n}{(2\pi)^4} e^{-i(q_1+\cdots +q_n)\cdot x}
(-1)\int\frac{d^4 k}{(2\pi)^4} tr[\sigma^{\mu\nu}\frac{i}{\not\! k-m}\gamma^{\rho_1}
\frac{i}{\not\! k-{\not\! q}_1-m}\cdots \nonumber \\
\times \gamma^{\rho_n}\frac{i}{\not\! k-{\not\! q}_1-\cdots -{\not\! q}_n-m}](ie)^n
{\tilde A}_{\rho_1}(q_1)\cdots {\tilde A}_{\rho_n}(q_n)
\end{eqnarray}
where ${\tilde A}_\rho(q)$ is the Fourier transform of $A_\rho(x)$: $A_\rho(x)=
\int\frac{d^4 q}{(2\pi)^4} e^{-iq \cdot x}{\tilde A}_\rho(q)$. Therefore 
the $A^n$ order contribution to $\partial_\mu\langle {\bar \psi}(x)\sigma^{\mu\nu}
\psi(x)\rangle$ is given by 
\begin{eqnarray}
\int \frac{d^4 q_1}{(2\pi)^4}\cdots \frac{d^4 q_n}{(2\pi)^4} e^{-i(q_1+\cdots +q_n)
\cdot x} i(q_1+\cdots +q_n)_\mu\int \frac{d^4 k}{(2\pi)^4} tr[\sigma^{\mu\nu}\frac{i}
{\not\! k-m}\gamma^{\rho_1}\frac{i}{\not\! k-{\not\! q}_1-m} \nonumber \\
\cdots \gamma^{\rho_n}\frac{i}{\not\! k-{\not\! q}_1-\cdots -{\not\! q}_n-m}]
(ie)^n {\tilde A}_{\rho_1}(q_1)\cdots {\tilde A}_{\rho_n}(q_n) 
\label{lhstotal}
\end{eqnarray}
Because the tensor current ${\bar \psi}(x)\sigma^{\mu\nu}\psi(x)$ is odd under 
charge conjugation, (\ref{lhstotal}) is nonvanishing only for odd $n$. For low
$n$, the loop integral in (\ref{lhstotal}) is divergent and needs to be regularized.
Here we choose dimensional regularization. The dimensionally regularized form 
of the loop integral in (\ref{lhstotal}) is 
\begin{equation}
i(q_1+\cdots +q_n)_\mu \int \frac{d^D k}{(2\pi)^D} tr[\sigma^{\mu\nu}\frac{i}{\not\! k-m}
\gamma^{\rho_1}\frac{i}{\not\! k-{\not\! q}_1-m}\cdots \gamma^{\rho_n}\frac{i}
{\not\! k-{\not\! q}_1-\cdots -{\not\! q}_n-m}] \label{lhsloop}
\end{equation}
Here all the indices and external momenta live in the physical four dimensions while
the loop momentum $k$ lives in $D$ dimensions. (\ref{lhsloop}) can be easily rewritten as
\begin{eqnarray}
-\frac{1}{2}\int \frac{d^D k}{(2\pi)^D} tr[({\not\! q}_1+\cdots +{\not\! q}_n)\gamma^{\nu}
\frac{i}{\not\! k-m}\gamma^{\rho_1}\frac{i}{\not\! k-{\not\! q}_1-m}\cdots\gamma^{\rho_n}
\frac{i}{\not\! k-{\not\! q}_1-\cdots -{\not\! q}_n-m}] \nonumber \\
+\frac{1}{2}\int \frac{d^D k}{(2\pi)^D} tr[\gamma^{\nu}({\not\! q}_1+\cdots +{\not\! q}_n)
\frac{i}{\not\! k-m}\gamma^{\rho_1}\frac{i}{\not\! k-{\not\! q}_1-m}\cdots\gamma^{\rho_n}
\frac{i}{\not\! k-{\not\! q}_1-\cdots -{\not\! q}_n-m}] \label{lhsfinal}
\end{eqnarray}

The $A^n$ order contribution to $-2m\langle {\bar \psi}(x)\gamma^{\nu}\psi(x)\rangle$ 
is also represented by an $(n+1)$-gon diagram (see fig.1) and equals
\begin{eqnarray}
\int\frac{d^4 q_1}{(2\pi)^4}\cdots \frac{d^4 q_n}{(2\pi)^4} e^{-i(q_1+\cdots +q_n)
\cdot x} 2m \int \frac{d^4 k}{(2\pi)^4} tr[\gamma^{\nu}\frac{i}{\not\! k-m}
\gamma^{\rho_1}\frac{i}{\not\! k-{\not\! q}_1-m} \cdots \nonumber \\
\times \gamma^{\rho_n}\frac{i}{\not\! k-{\not\! q}_1-\cdots -{\not\! q}_n-m}]
(ie)^n {\tilde A}_{\rho_1}(q_1)\cdots {\tilde A}_{\rho_n}(q_n) \label{rhs1total}
\end{eqnarray}
Because the vector current ${\bar \psi}(x)\gamma^{\nu}\psi(x)$ is odd under
charge conjugation, (\ref{rhs1total}) is nonvanishing only for odd $n$. 
The dimensionally regularized form of the loop integral in (\ref{rhs1total})
is 
\begin{equation}
2m\int\frac{d^D k}{(2\pi)^D} tr[\gamma^{\nu}\frac{i}{\not\! k-m}\gamma^{\rho_1}
\frac{i}{\not\! k-{\not\! q}_1-m}\cdots \gamma^{\rho_n}\frac{i}{\not\! k-
{\not\! q}_1-\cdots - {\not\! q}_n-m}] \label{rhs1final}
\end{equation}
  
The $A^n$ order contribution to $i\langle {\bar \psi}(x)(\stackrel{\rightarrow}
{\partial^{\nu}}-\stackrel{\leftarrow}{\partial^{\nu}})\psi(x)\rangle$ is also 
represented by an $(n+1)$-gon diagram (see fig.1) and equals 
\begin{eqnarray}
&&i\int \frac{d^4 q_1}{(2\pi)^4}\cdots \frac{d^4 q_n}{(2\pi)^4} e^{-i(q_1+\cdots
+ q_n) \cdot x}(-1)\int \frac{d^4 k}{(2\pi)^4} tr[\frac{i}{\not\! k-m}\gamma^
{\rho_1}\frac{i}{\not\! k-{\not\! q}_1-m}\cdots \gamma^{\rho_n} \nonumber \\
&&\times \frac{i}{\not\! k-{\not\! q}_1 -\cdots - {\not\! q}_n-m}]((-i)k^{\nu}
-i(k-q_1-\cdots -q_n)^{\nu}) (ie)^n {\tilde A}_{\rho_1}(q_1)\cdots 
{\tilde A}_{\rho_n}(q_n) \nonumber \\
&=& -\int \frac{d^4 q_1}{(2\pi)^4}\cdots\frac{d^4 q_n}{(2\pi)^4} e^{-i(q_1+\cdots
+q_n)\cdot x}\int \frac{d^4 k}{(2\pi)^4} tr[\frac{i}{\not\! k-m}\gamma^{\rho_1}
\frac{i}{\not\! k-{\not\! q}_1-m}\cdots \gamma^{\rho_n} \nonumber \\
&&\times \frac{i}{\not\! k-{\not\! q}_1-\cdots -{\not\! q}_n-m}]
(2k-q_1-\cdots -q_n)^{\nu}(ie)^n{\tilde A}_{\rho_1}(q_1)\cdots{\tilde A}_{\rho_n}
(q_n) \label{rhs2total}
\end{eqnarray}
Because the operator ${\bar \psi}(x)(\stackrel{\rightarrow}{\partial^{\nu}}-
\stackrel{\leftarrow}{\partial^{\nu}})\psi(x)$ is odd under charge conjugation,
(\ref{rhs2total}) is nonvanishing only for odd $n$. The dimensionally regularized 
form of the loop integral in (\ref{rhs2total}) is 
\begin{equation}
-\int \frac{d^D k}{(2\pi)^D} tr[\frac{i}{\not\! k-m}\gamma^{\rho_1}\frac{i}
{\not\! k-{\not\! q}_1-m}\cdots \gamma^{\rho_n}\frac{i}{\not\! k-{\not\! q}_1-
\cdots-{\not\! q}_n-m}](2k-q_1-\cdots -q_n)^{\nu} \label{rhs2loop}
\end{equation}
Now we write
\begin{eqnarray}
&&-\int \frac{d^D k}{(2\pi)^D} tr[\frac{i}{\not\! k-m}\gamma^{\rho_1}\frac{i}
{\not\! k-{\not\! q}_1-m}\cdots \gamma^{\rho_n}\frac{i}{\not\! k-{\not\! q}_1-
\cdots-{\not\! q}_n-m}](2k-q_1-\cdots -q_n)^{\nu} \nonumber \\
&=&-\frac{1}{2}\int \frac{d^D k}{(2\pi)^D} tr[\frac{i}{\not\! k-m}\gamma^{\rho_1}\frac{i}
{\not\! k-{\not\! q}_1-m}\cdots \gamma^{\rho_n}\frac{i}{\not\! k-{\not\! q}_1-
\cdots-{\not\! q}_n-m}\{2\not\! k-{\not\! q}_1-\cdots -{\not\! q}_n,\gamma^{\nu}\}]
\nonumber \\
&=& -\frac{1}{2}\int \frac{d^D k}{(2\pi)^D} tr[\frac{i}{\not\! k-m}\gamma^{\rho_1}\frac{i}
{\not\! k-{\not\! q}_1-m}\cdots \gamma^{\rho_n}\frac{i}{\not\! k-{\not\! q}_1-
\cdots-{\not\! q}_n-m}\{2\not\! k-{\not\! q}_1-\cdots -{\not\! q}_n-2m,\gamma^{\nu}\}]
\nonumber \\
&&-\frac{1}{2}\int \frac{d^D k}{(2\pi)^D} tr[\frac{i}{\not\! k-m}\gamma^{\rho_1}\frac{i}
{\not\! k-{\not\! q}_1-m}\cdots \gamma^{\rho_n}\frac{i}{\not\! k-{\not\! q}_1-
\cdots-{\not\! q}_n-m}\{2m,\gamma^{\nu}\}] \nonumber \\
&=& -\int \frac{d^D k}{(2\pi)^D} tr[\frac{i}{\not\! k-m}\gamma^{\rho_1}\frac{i}
{\not\! k-{\not\! q}_1-m}\cdots \gamma^{\rho_n}\frac{i}{\not\! k-{\not\! q}_1-
\cdots-{\not\! q}_n-m}\{\not\! k-m,\gamma^{\nu}\}] \nonumber \\
&&+\frac{1}{2}\int \frac{d^D k}{(2\pi)^D} tr[\frac{i}{\not\! k-m}\gamma^{\rho_1}\frac{i}
{\not\! k-{\not\! q}_1-m}\cdots \gamma^{\rho_n}\frac{i}{\not\! k-{\not\! q}_1-
\cdots-{\not\! q}_n-m}\{{\not\! q}_1+\cdots +{\not\! q}_n,\gamma^{\nu}\}] \nonumber \\
&&-2m \int\frac{d^D k}{(2\pi)^D} tr[\gamma^{\nu}\frac{i}{\not\! k-m}\gamma^{\rho_1}\frac{i}
{\not\! k-{\not\! q}_1-m}\cdots \gamma^{\rho_n}\frac{i}{\not\! k-{\not\! q}_1-
\cdots-{\not\! q}_n-m}] \label{rhs2final}
\end{eqnarray}

To calculate the $A^n$ order contribution to $2e\langle{\bar \psi}(x)\psi(x)\rangle
A^{\nu}(x)$, one needs to calculate the $A^{n-1}$ order contribution to $\langle
{\bar \psi}(x)\psi(x)\rangle$. The latter is represented by an $n$-gon diagram (see
fig.2) and equals
\begin{eqnarray}
\int \frac{d^4 q_1}{(2\pi)^4} \cdots \frac{d^4 q_{n-1}}{(2\pi)^4} e^{-i(q_1+\cdots+
q_{n-1})\cdot x}(-1)\int \frac{d^4 k}{(2\pi)^4} tr[\frac{i}{\not\! k-m}\gamma^{\rho_1}
\frac{i}{\not\! k-{\not\! q}_1-m}\cdots \gamma^{\rho_{n-1}} \nonumber \\
\times \frac{i}{\not\! k-{\not\! q}_1-\cdots -{\not\! q}_{n-1}-m}](ie)^{n-1}
{\tilde A}_{\rho_1}(q_1)\cdots {\tilde A}_{\rho_{n-1}}(q_{n-1}) \label{rhs3}
\end{eqnarray}
After multiplying by $2e A^{\nu}(x) \rightarrow 2e g^{\nu\rho_n}A_{\rho_n}(x)=
2e g^{\nu\rho_n}\int \frac{d^4 q_n}{(2\pi)^4} e^{-iq_n \cdot x} {\tilde A}_{\rho_n}
(q_n)$ we get the $A^n$ order contribution to $2e\langle {\bar \psi}(x)\psi(x)\rangle
A^{\nu}(x)$:
\begin{eqnarray}
\int \frac{d^4 q_1}{(2\pi)^4}\cdots \frac{d^4 q_n}{(2\pi)^4} e^{-i(q_1+\cdots+
q_n)\cdot x}(2i)\int \frac{d^4 k}{(2\pi)^4} tr[\frac{i}{\not\! k-m}\gamma^{\rho_1}
\frac{i}{\not\! k-{\not\! q}_1-m}\cdots \gamma^{\rho_{n-1}} \nonumber \\
\times\frac{i}{\not\! k-{\not\! q}_1-\cdots -{\not\! q}_{n-1}-m}]g^{\nu\rho_n}(ie)^n
{\tilde A}_{\rho_1}(q_1)\cdots{\tilde A}_{\rho_n}(q_n) \label{rhs3total}
\end{eqnarray}
Because the scalar operator ${\bar \psi}(x)\psi(x)$ is even under charge conjugation,
(\ref{rhs3})(and also (\ref{rhs3total}))is nonvanishing only for odd $n$. The 
dimensionally regularized form of the loop integral in (\ref{rhs3total}) is
\begin{equation}
2i\int \frac{d^D k}{(2\pi)^D} tr[\frac{i}{\not\! k-m}\gamma^{\rho_1}\frac{i}
{\not\! k-{\not\! q}_1-m}\cdots \gamma^{\rho_{n-1}}\frac{i}{\not\! k-{\not\! q}_1
-\cdots -{\not\! q}_{n-1}-m}]g^{\nu\rho_n} \label{rhs3loop}
\end{equation}
Now we write
\begin{eqnarray}
&&2i\int \frac{d^D k}{(2\pi)^D} tr[\frac{i}{\not\! k-m}\gamma^{\rho_1}\frac{i}
{\not\! k-{\not\! q}_1-m}\cdots \gamma^{\rho_{n-1}}\frac{i}{\not\! k-{\not\! q}_1
-\cdots -{\not\! q}_{n-1}-m}]g^{\nu\rho_n} \nonumber \\
&=& i\int\frac{d^D k}{(2\pi)^D} tr[\frac{i}{\not\! k-m}\gamma^{\rho_1}\frac{i}
{\not\! k-{\not\! q}_1-m}\cdots \gamma^{\rho_{n-1}}\frac{i}{\not\! k-{\not\! q}_1
-\cdots -{\not\! q}_{n-1}-m}\{\gamma^{\rho_n},\gamma^{\nu}\}] \nonumber \\
&=& i\int\frac{d^D k}{(2\pi)^D} tr[\frac{i}{\not\! k-m}\gamma^{\rho_1}\frac{i}
{\not\! k-{\not\! q}_1-m}\cdots \gamma^{\rho_{n-1}}\frac{i}{\not\! k-{\not\! q}_1
-\cdots -{\not\! q}_{n-1}-m}\gamma^{\rho_n}\gamma^{\nu}] \nonumber \\
&&+i\int \frac{d^D k}{(2\pi)^D} tr[\gamma^{\rho_n}\frac{i}{\not\! k-m}\gamma^{\rho_1}\frac{i}
{\not\! k-{\not\! q}_1-m}\cdots \gamma^{\rho_{n-1}}\frac{i}{\not\! k-{\not\! q}_1
-\cdots -{\not\! q}_{n-1}-m}\gamma^{\nu}] 
\end{eqnarray}
In the second term of the last line, we rename the integration(summation) variables
(indices): $(q_n,\rho_n) \rightarrow (q_1,\rho_1),(q_1,\rho_1)\rightarrow (q_2,\rho_2),
\cdots (q_{n-1},\rho_{n-1}) \rightarrow (q_n,\rho_n)$(this renaming does not alter 
the value of $\int \frac{d^4 q_1}{(2\pi)^4}\cdots\frac{d^4 q_n}{(2\pi)^4}\cdots
{\tilde A}_{\rho_1}(q_1)\cdots {\tilde A}_{\rho_n}(q_n)$) and obtain
\begin{equation}
i\int\frac{d^D k}{(2\pi)^D} tr[\gamma^{\rho_1}\frac{i}{\not\! k-m}\gamma^{\rho_2}
\frac{i}{\not\! k-{\not\! q}_2-m}\cdots \gamma^{\rho_n}\frac{i}{\not\! k-{\not\! q}_2
-\cdots -{\not\! q}_n-m}\gamma^{\nu}]
\end{equation}
After making the shift: $k \rightarrow k-q_1$, we get
\begin{equation}
i\int\frac{d^D k}{(2\pi)^D} tr[\gamma^{\rho_1}\frac{i}{\not\! k-{\not\! q}_1-m}
\gamma^{\rho_2}\frac{i}{\not\! k-{\not\! q}_1-{\not\! q}_2-m}\cdots \gamma^{\rho_n}
\frac{i}{\not\! k-{\not\! q}_1-{\not\! q}_2-\cdots -{\not\! q}_n-m}\gamma^{\nu}]
\end{equation}
So we can replace (\ref{rhs3loop}) by 
\begin{eqnarray}
&&i\int \frac{d^D k}{(2\pi)^D} tr[\frac{i}{\not\! k-m}\gamma^{\rho_1}\frac{i}
{\not\! k-{\not\! q}_1-m}\cdots \gamma^{\rho_n}\gamma^{\nu}] \nonumber \\
&&+i\int\frac{d^D k}{(2\pi)^D}tr[\gamma^{\rho_1}\frac{i}{\not\! k-{\not\! q}_1-m}
\cdots \gamma^{\rho_n}\frac{i}{\not\! k-{\not\! q}_1-\cdots -{\not\! q}_n-m}
\gamma^{\nu}]
\end{eqnarray}
Now we do further reductions:
\begin{eqnarray}
&&i\int \frac{d^D k}{(2\pi)^D} tr[\frac{i}{\not\! k-m}\gamma^{\rho_1}\frac{i}
{\not\! k-{\not\! q}_1-m}\cdots \gamma^{\rho_n}\gamma^{\nu}] \nonumber \\
&&+i\int\frac{d^D k}{(2\pi)^D}tr[\gamma^{\rho_1}\frac{i}{\not\! k-{\not\! q}_1-m}
\cdots \gamma^{\rho_n}\frac{i}{\not\! k-{\not\! q}_1-\cdots -{\not\! q}_n-m}
\gamma^{\nu}] \nonumber \\
&=& i\int \frac{d^D k}{(2\pi)^D}tr[\frac{i}{\not\! k-m}\gamma^{\rho_1}
\frac{i}{\not\! k-{\not\! q}_1-m}\cdots \gamma^{\rho_n}\frac{i}{\not\! k-
{\not\! q}_1-\cdots-{\not\! q}_n-m}(-i)(\not\! k-{\not\! q}_1-\cdots-
{\not\! q}_n-m)\gamma^{\nu}] \nonumber \\
&&+i\int\frac{d^D k}{(2\pi)^D}tr[(-i)(\not\! k-m)\frac{i}{\not\! k-m}\gamma^{\rho_1}
\frac{i}{\not\! k-{\not\! q}_1-m}\cdots \gamma^{\rho_n}\frac{i}{\not\! k-
{\not\! q}_1-\cdots -{\not\! q}_n-m}\gamma^{\nu}] \nonumber \\
&=& \int \frac{d^D k}{(2\pi)^D} tr[\frac{i}{\not\! k-m}\gamma^{\rho_1}
\frac{i}{\not\! k-{\not\! q}_1-m}\cdots \gamma^{\rho_n}\frac{i}
{\not\! k-{\not\! q}_1-\cdots -{\not\! q}_n-m}(\not\! k-m)\gamma^{\nu}] \nonumber \\
&&-\int \frac{d^D k}{(2\pi)^D} tr[\frac{i}{\not\! k-m}\gamma^{\rho_1}
\frac{i}{\not\! k-{\not\! q}_1-m}\cdots \gamma^{\rho_n}\frac{i}
{\not\! k-{\not\! q}_1-\cdots -{\not\! q}_n-m}({\not\! q}_1+\cdots +
{\not\! q}_n)\gamma^{\nu}] \nonumber \\
&&+\int \frac{d^D k}{(2\pi)^D} tr[\frac{i}{\not\! k-m}\gamma^{\rho_1}
\frac{i}{\not\! k-{\not\! q}_1-m}\cdots \gamma^{\rho_n}\frac{i}
{\not\! k-{\not\! q}_1-\cdots -{\not\! q}_n-m}\gamma^{\nu}(\not\! k-m)]
\label{rhs3final}
\end{eqnarray}
Comparing (\ref{lhsfinal}),(\ref{rhs1final}),(\ref{rhs2final}) and
(\ref{rhs3final}) we find that the divergence equation of the tensor
current is satisfied (at the level of VEVs) and is therefore free
of anomalies.

To summarize, in this paper we show by explicit perturbative calculations
that the divergence equation of the tensor current in $U(1)$ gauge theories
is free of anomalies.
 
This work is supported in part by the National Natural Science Foundation
of China under Grant No. 90103018, 10175033 and 10135030.

\begin{figure}
\includegraphics{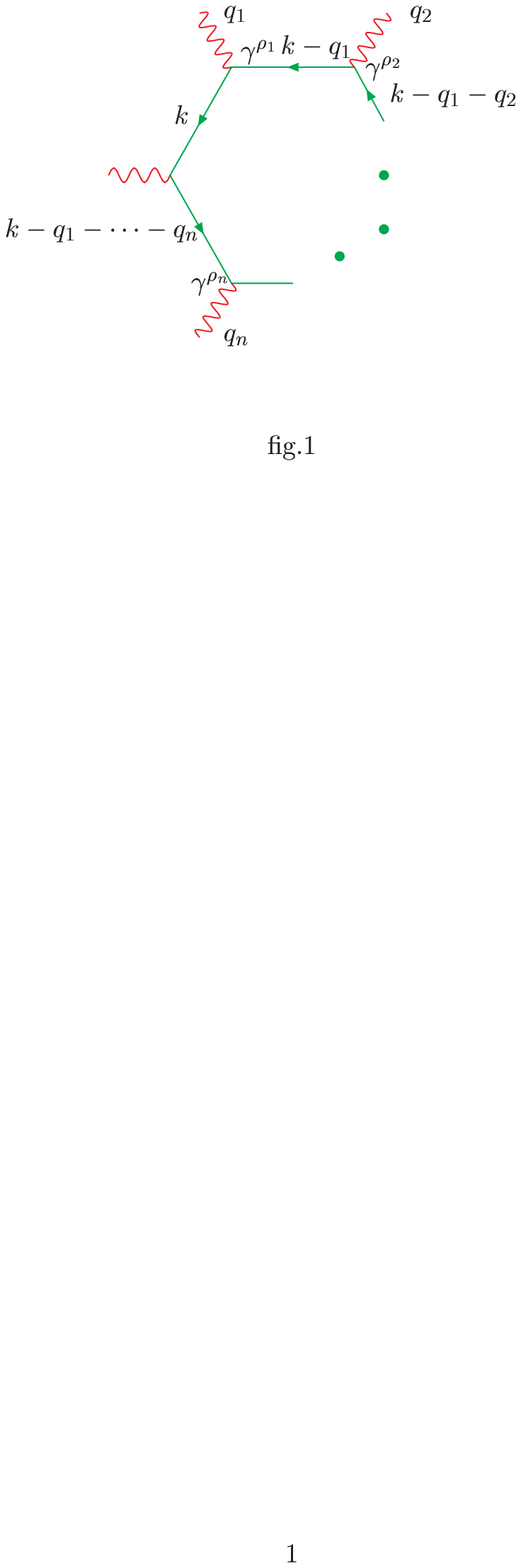}
\end{figure}

\begin{figure}
\includegraphics{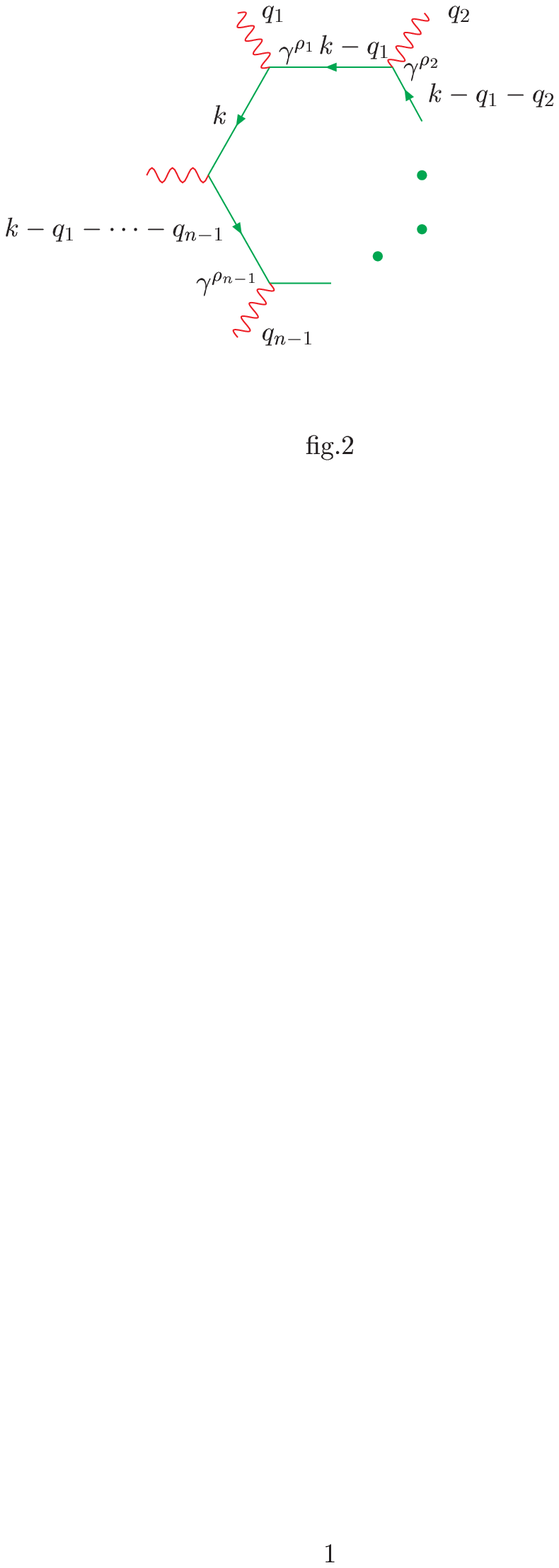}
\end{figure}

\end{document}